\documentclass[draftclsnofoot,onecolumn, a4paper,11pt]{IEEEtran}

\usepackage{epsfig,amssymb,amsmath}        
\usepackage{algorithmic}
\usepackage{algorithm}

\usepackage{subfigure}

\newcounter{theorem}
\newtheorem{theorem}{\noindent\textbf{\emph{Theorem}}}

\newcounter{definition}

\newcounter{claim} 

\newcounter{lemma} \newtheorem{lemma}{\noindent\textbf{\emph{Lemma}}}
\newcounter{proposition} 

\begin{document}
\bibliographystyle{plain}
%
\title{ReCord: A Distributed Hash Table with Recursive Structure}
%
%


\author{Jianyang~Zeng and Wen-Jing~Hsu}



\maketitle

\begin{abstract}
We propose a simple distributed hash table called ReCord, which is
a generalized version of Randomized-Chord and offers improved
tradeoffs in performance and topology maintenance over existing
P2P systems. ReCord is scalable and can be easily implemented as
an overlay network, and offers a good tradeoff between the node
degree and query latency. For instance, an $n$-node ReCord with
$O(\log  n)$ node degree has an expected latency of $\Theta(\log
n)$ hops. Alternatively, it can also offer $\Theta(\frac{\log
n}{\log \log
 n})$ hops latency at a higher cost of $O(\frac{\log^2  n}{\log
 \log n})$ node degree. Meanwhile, simulations of the dynamic behaviors
of ReCord are studied.
\end{abstract}


\noindent \textbf{Corresponding authors:}

\noindent (1). Dr. Wen-Jing Hsu \qquad Tel: (065)67904597   \qquad E-mail: hsu@ntu.edu.sg; \\
(2). Mr. Jianyang Zeng \qquad Tel: (065)67906333 \qquad E-mail:
pg03858494@ntu.edu.sg

 \noindent \textbf{Paper type:} regular

\noindent \textbf{Affiliation and Postal address:}

\noindent Centre for Advanced Information Systems, School of
Computer Engineering, Nanyang Technological University, Nanyang
Avenue, Singapore 639798

\IEEEpeerreviewmaketitle


%


\section{Introduction}
Peer to peer (P2P) networks have become popular in resource
sharing applications recently. There have been millions of users
in certain successful systems, such as Napster~\cite{Napster},
Gnutella~\cite{Gnu}, and Kazaa~\cite{Kazaa}. P2P systems are
distributed systems without any central authority or hierarchical
organization, and each node in the systems performs similar
functionality.

In order to efficiently locate an object in a large scale P2P
system, many schemes relay on distributed hash tables (DHTs).
Example systems include Chord~\cite{SMKKB01}, Pastry~\cite{RD01},
CAN~\cite{RFHK01}, and Tapestry~\cite{ZKJ01}. A P2P system needs
to consider the joining, departing of hosts, and the
insertion/addition of resources, besides look up operation. DHT
can be implemented as an overlay logical topology over the
internet physical networks, where each node keeps the direct IP
address of its neighbors in a routing table. Instead of connecting
to all the other nodes in the system, each node in DHT only links
to a small number of nodes. The key lies in ensuring a small
\emph{diameter} in the resulting overlay network. At the same
time, DHT should allow new nodes to join or existing nodes to
leave the system voluntarily, therefore, the cost of topology
maintenance during this dynamic join-depart process should be kept
as low as possible.

The following metrics are usually used to compare the performance
and efficiency of the designed DHT.

\begin{enumerate}
\item[(a)]\textbf{Degree and Diameter}: The number of links per
node in a DHT should be small in order to reduce the cost of
communication. Also, the diameter of the network should not be
large in order to reduce the query latency.

\item[(b)]\textbf{Scalability}: As the network size increases, the
node degree, the query latency, and the traffic increased in
queries should not increase drastically.

\item[(c)] \textbf{Maintenance overhead}: When new nodes join or
existing nodes depart, the overhead as measured in terms of the
number of messages required to maintain the DHT should be as low
as possible.

\item[(d)] \textbf{Fault tolerance}: The DHT should be resilient
to both node and link failures of the system. No matter what
fraction of the nodes or links has failed, the data available in
the remaining nodes should still be accessible.

\item[(e)] \textbf{Load balance}: The resource keys should be
evenly distributed over all nodes, and the traffic overhead
resulted by query or maintenance operations should be balanced
among nodes in the network.
\end{enumerate}

In this paper, we propose a simple distributed hash table, called
ReCord, which is scalable and can be easily implemented as an
overlay network. ReCord offers a good tradeoff between the node
degree and query latency. For instance, an $n$-node ReCord with
$O(\log  n)$ node degree has an expected latency of $\Theta(\log
 n)$ hops. Alternatively, it can also offer  $\Theta(\frac{\log
n}{ \log \log n})$  hops latency at a higher cost of
$O(\frac{\log^2
 n}{\log \log  n})$ node degree.

The rest of the paper is organized as follows. Section~2 will
review related work. In Section~3, the construction of ReCord will
be described. Section~4 will examine the bounds of the node degree
and route path length. Section~5 presents an implementation of
ReCord. Section~6 gives the simulation studies of ReCord's dynamic
behavors. Section~7 summarizes the finding and concludes this
paper.

\section{Related work}

Plaxton~\cite{PRR97} \textit{et al.} proposed a distributed
routing protocol based on hypercubes for a static network with
given collection of nodes. Plaxton's algorithm uses the
\emph{digit-fixing} technique to locate the shared resources on an
overlay network in which each node only maintains a small-sized
routing table. Pastry~\cite{RD01} and Tapestry~\cite{ZKJ01} use
Plaxton's scheme in the dynamic distributed environment. The
difference between them is that Pastry uses \emph{prefix-based}
routing scheme, whereas Tapestry uses \emph{suffix-based} scheme.
The number of bits per digit for both Tapestry and Pastry can be
reconfigured but it remains fixed during run-time. Both Pastry and
Tapestry can build the overlay topology using proximity neighbor
selection. However, it is still unclear whether there is any
better approach to achieve globally effective routing.

Chord~\cite{SMKKB01} uses consistent hashing method to map $n$
nodes to evenly distribute around a circle of identifiers. Each
node $x$ in Chord stores a pointer to its immediate
\emph{successor} (the closest node in clockwise direction along
the circle), and a \emph{finger table} of connections with node
$x+2^{i}$, where $i=1,2,{...}, \log  n-1$. In Chord, a greedy
algorithm is used to route query messages.  The complexity of
routing per query is bounded by $O(\log n)$ hops. For fault
tolerance, each node in Chord uses a successor list which stores
the connections of next several successor nodes. The routing
protocol in standard Chord in~\cite{SMKKB01} is not optimal and
was improved by using bidirectional
connections~\cite{GM04optimal}. In~\cite{EAB02}, EI-Ansary et al.
generalize Chord to a P2P framework with $k$-ary search, but they
only focused on  the lookup operation, and did not consider node
joining leaving, failure, and implementation details.
In~\cite{GGG03impact,MNW04}, a randomized version of Chord, called
\emph{Randomized-Chord}, is presented. In Randomized-Chord, node
$s$ is connected to a randomly chosen node in each interval
$[2^{i-1},2^i)$, where $i=1,2,{...}, \log  n$. Koorde~\cite{KK03}
is an extended DHT of Chord in that it embeds a de Bruijn graph
over the Chord identifier circle. Koorde can be constructed with
constant node degree and $O(\log n)$ hops per query, or with
$O(\log  n)$ node degree and $O( \log n/ \log \log n)$ hops per
query. As a Chord-like network, Symphony~\cite{MBR03} builds a
network using the \emph{small world} model
from~\cite{Kle00,BFK01}. In Symphony, each node has local links
with its immediate neighbors and long distance links connected to
randomly chosen nodes from a probability distribution function.
The expected path length for a Symphony network with $k$ links is
$O(\frac{1}{k} \log^2  n)$ hops. Simulations in~\cite{MBR03} shows
that Symphony is scalable, flexible, stable in the dynamic
environment, and offers a small average latency with constant
degree, but the analytical results for fault tolerance were not
given. Like Chord, Viceroy~\cite{MNR02} distributes nodes along a
circle, and builds a constant-degree topology approximating a
butterfly network, and offers $O(\log  n)$ routing latency.
However, it is relatively complicated to implement Viceroy and
fault tolerance is not addressed in~\cite{MNR02}.

CAN~\cite{RFHK01} divides a $d$-dimension torus space into zones
\emph{owned} by nodes, and resource keys are evenly hashed into
the coordinate space. Each resource key is stored at the node that
owns the located zone. Using greedy routing, the query message is
routed to the neighbor which is closer to the target key. Each
node has $O(d)$ neighbors and query latency is $O(dn^{1/d})$. If
$d$ is chosen to be $\log  n$, each node connects with $O(\log n)$
neighbors and a query takes $O(\log  n)$ hops. Some proximity
routing scheme, such as \emph{global network
positioning}~\cite{NZ01towards} and topologically-aware overlay
construction~\cite{RHS02topologically} to build CAN overlay
network. There are two disadvantages for this scheme: it needs to
fix some landmark machines and it tends to create hot spots from a
non-uniform distribution of nodes in the coordinate space.

It is difficult to say which one of above proposed DHTs is
``best".  Each routing algorithm offers some insight on routing in
overlay network. One appropriate strategy is to combine these
insights and formulate an even better scheme~\cite{RST02routing}.

\section{Construction of ReCord} \label{sec:construction}
In this paper, we will slightly abuse the notation of node
identifiers and nodes themselves, and the same to resource key
identifiers and resource themselves. Instead of mapping
identifiers into $m$-bit numbers, we will map them into the unit
circle ring $\mathcal{I} \ [0,1)$, as with Symphony~\cite{MBR03}
and Viceroy~\cite{MNR02}. By using a consistent hashing method, we
can assume that both node and key identifiers are distributed
evenly over the circle $[0,1)$, and there is no collision.

Hashing the identifiers into ring $\mathcal{I} \  [0,1)$ allows
the identifier value to be independent of the maximum hashed space
$2^m$.
Assume that the ring is formed in the clockwise direction. Denote
the clockwise neighbor of node $s$ on the ring by $SUCC(s)$, and
denote its counter-clockwise neighbor by $PRED(s)$. A key $x$ is
stored at a nearest node $y$, where $y \geq x$ on the ring. We
also call this node $SUCC(x)$.

\begin{figure}
\centering
\includegraphics[width=4in]{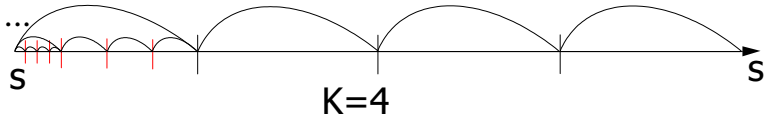}
\caption{An example of interval division ($k=4$).}
\label{fig:divisionexample}
\end{figure}

The basic idea of ReCord is as follows. Suppose that there are
totally $n$ active nodes in a stable P2P system. Starting from any
node $s$, divide the whole ring into $k$ equal intervals, where
$k>1$ denotes an integer. Then divide the first interval closest
to node $s$ recursively until the length of the interval nearest
to $s$ is $\frac{1}{n}$, i.e. the first $k$ intervals nearest to
$s$ contains $O(1)$ nodes, the second $k$ intervals nearest to $s$
contains $O(k)$ nodes, and the third $O(k^2)$ nodes and so on, as
shown in Fig.~\ref{fig:divisionexample}.

The first division is also called level-1 division, and the next
is called level-2 division, and so on. There are $c= \log_k \ n$
such levels (assuming that $n=k^c$, where $c$ denotes an integer).
The length of each interval at level 1 is $\frac{1}{k}$, and
$\frac{1}{k^2}$ for level 2, and $\frac{1}{k^i}$ at level $i$ in
general. The intervals at the same level are numbered sequentially
clockwise along the ring. There are totally $k$ intervals in every
level. Based on the above definitions, for node $s$, we know that
its interval $j$ at level $i$ corresponds to the range
$[s+\frac{(j-1)k^i}{n},s+\frac{jk^i}{n})$ on the ring. Randomly
choose one node $x$ in every interval, and set up a unidirectional
connection from node $s$ to $x$. We call the resulting network
`ReCord' for its recursive structure and similarity with Chord.

Comparing ReCord with Randomized-Chord, we find that in fact
ReCord is a generalized version of Randomized-Chord. When $k=2$,
ReCord becomes Randomized-Chord.

\section{Analysis} \label{sec:bound_analysis}

P2P systems have dynamic membership, meaning that a node may join
and leave the network voluntarily. The number of the active nodes
varies with the evolution of the P2P network. However, when we
analyze the degree of each node or latency of each query, we
suppose that the P2P network is static. Therefore, we will firstly
analyze the tradeoffs between link degree and query latency for
ReCord statically. Later, we will explain how to extend and
implement it under the dynamic situation.

\begin{theorem}
The node degree per node in an $n$-node ReCord is $\Theta(k \log_k
n)$.
\end{theorem}
\noindent \textbf{Proof:}\quad Let $H(n)$ represent the number of
links connected to an arbitrary node $s$ in the $n$-node network.
After the first division, there are $k-1$ links, plus its links to
nodes in the intervals included by level-2, hence we have the
following relation:
\[
H(n)=(k-1)+H(\frac{n}{k}).
\]

The solution of this recurrence is $H(n)=\Theta \big((k-1) \log_k
n \big)=\Theta(k \log_k  n)$. Therefore, the degree of any node in
ReCord is bounded by $\Theta(k \log_k  n)$.~\
\hfill\rule{2.0mm}{2.6mm}

When $k=\Theta (1)$, the node degree in the $n$-node ReCord is
$H(n)=\Theta(\log n)$. If $k=\Theta (\log  n)$,
$H(n)=\Theta(\frac{\log^2 n }{\log \log  n})$.

Before studying the query latency in ReCord, we introduce the
following lemma which will be used in the proof of the lower bound
of the query latency.

\begin{lemma} \label{lemma:lowerbound}
Let $X^m$ denote a random variable in the state space $0,1,2,\cdot
\cdot \cdot m-1$. Let
\[
Pr[X^m=i]=
 \left\{
 \begin{array}{cl}
\frac{k-1}{k^{m-i}},& \mbox{when $1 \leq i \leq m-1$} \\
\frac{1}{k^{m}},& \mbox{when $i =0$}
\end{array}
   \right.
\]
The expected time required for $X^m$ to drop to 0 is lower bounded
by $E[T_m]=\Omega (m)$
\end{lemma}

\noindent \textbf{Proof:}\quad We apply an inductive method in
this proof. The lemma holds obviously when $i=0,1$. Suppose that
the lemma holds for all $1 < i \leq m-1$, namely, $E[T_i] \geq c
i$, for all $1 < i \leq m-1$, where $c$ is
a constant. We have\\
\noindent $ E[T_m] \geq 1+c(m-1)\frac{k-1}{k}+c(m-2)
\frac{k-1}{k^2}+\cdot \cdot \cdot +c\frac{k-1}{k^{m-1}} $

\noindent \ \ \ \ \ \ \ \ $
=1+cm(k-1)(\frac{1}{k}+\frac{1}{k^2}+\cdot \cdot \cdot
\frac{1}{k^{m-1}})-c(k-1)[\frac{1}{k}+2\frac{1}{k^2}+\cdot \cdot
\cdot +(m-1)\frac{1}{k^{m-1}}]$

\noindent \ \ \ \ \ \ \ \ $ \geq
1+cm(1-\frac{1}{k^m})-c(k-1)\frac{\frac{1}{k}}{(1-\frac{1}{k})^2}
$

\noindent \ \ \ \ \ \ \ \ $ =cm+1-\frac{cm}{k^m}-\frac{ck}{k-1} $.

When $c \leq \frac{1}{\frac{m}{k^m}+\frac{k}{k-1}}<1$, $E[T_m]
\geq cm$. Therefore, the expected time required for $X^m$ to drop
to 0 is lower bounded by $E[T_m]=\Omega (m)$, where the constant
multiplier in the formula is smaller than 1.~\
\hfill\rule{2.0mm}{2.6mm}

\begin{theorem}
Using the greedy routing algorithm, the expected path length per
query in an $n$-node ReCord is $\Theta(\log_k  n)$.
\end{theorem}
\noindent \textbf{Proof:}\quad \textbf{Upper bound:} Let $T(n)$
denote the number of hops required by a query. Consider the case
when the message is routed to the 1st interval, according to the
recursive construction, the time step required is
$T(n)=T(\frac{n}{k})$. If the message is routed to interval $j$ of
the level-division ($1<j \leq k$), in the worst case, the distance
is reduced to $\frac{2n}{k}-1$. In this case, after one more
forwarding, the distance will be reduced to less than
$\frac{n}{k}$, so the required time in the worst case is
upper-bounded by $T(n)\leq 2+T(\frac{n}{k})$. Since each link is
connected to a randomly chosen node in each interval, the
probability that the message is routed to interval 1 is
$\frac{k}{n}$, and the probability that it is routed to intervals
2,3,..., $k$ is $\frac{n-k}{n}$. Thus, an  upper bound of the
total expected number of steps is:
\begin{equation}
T(n) \leq
\frac{1}{k}T(\frac{k}{n})+\frac{k-1}{k}[2+T(\frac{k}{n})].
\label{equ:recursive_latency}
\end{equation}

Solving Ineq.~(\ref{equ:recursive_latency}), we have $T(n)=O \big(
\frac{2(k-1)}{k}  {\log}_k n \big)= O (\log_k n)$.  Therefore, for
the greedy routing algorithm, the expected path length per query
is $O(\log_k  n)$, where the constant multiplier in the formula is
smaller than 2.

\textbf{Lower bound:} Suppose that all nodes in the ring are
labelled by $0,1,2,\cdot \cdot \cdot ,n$.  Node 0 is the
destination, and node $n$ is the source. We define \emph{Phase} as
follows: Phase 0 only contains node 0; Phase 1 consists of nodes
in the interval $[1,k-1]$; Phase 2 consists of nodes in the
interval $[k,k^2-1]$, and so on. Generally, Phase $i$ contains
nodes in the interval $[k^{i-1},k^i-1]$. Suppose that there are in
total $m$ phases.

According to the division of intervals and randomly choosing one
node among each interval, the probability that the message is
routed to Phase $m-1$ is $\frac{k-1}{k}$, and $\frac{k-1}{k^2}$ if
routed to Phase $m-2$, and so forth. Generally, the probability
that the message is routed to Phase $i$  is $\frac{k-1}{k^{m-1}}$,
for $1\leq i \leq m-1$, and $\frac{1}{k^m}$, for $i=0$. By
applying Lemma~\ref{lemma:lowerbound}, we can deduce that the
expected number of hops per query is $\Omega(m)$. There are
totally $m= \log_k n$ phases for $n$ nodes. Therefore, the average
number of hops per query is lower bounded by $\Omega( \log_k
n)$.~\ \hfill\rule{2.0mm}{2.6mm}

\begin{figure}
\centering
\includegraphics[width=4in]{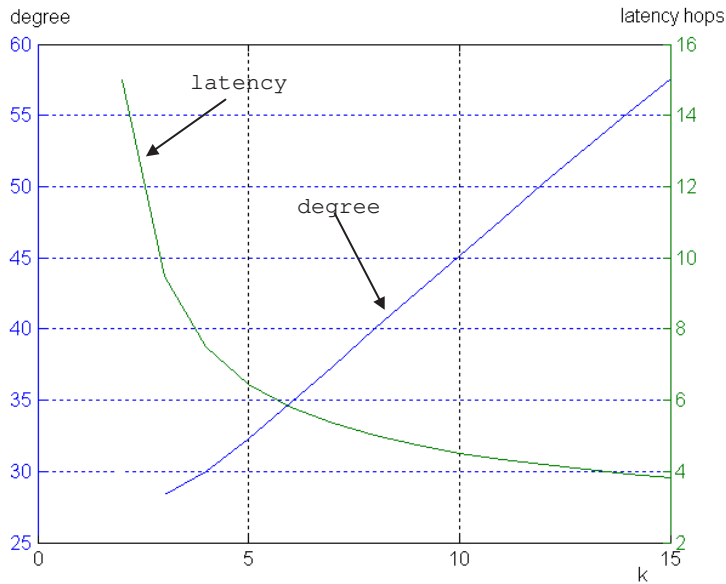}
\caption{Impact of variable $k$ on degree and latency.}
\label{fig:degree_latency}
\end{figure}

Our static analysis shows a good tradeoff between the node degree
and the required hops per query. If we choose $k=\Theta (1)$, the
node degree and query latency for an $n$-node ReCord are $O(log \
n)$ and $\Theta (log \ n)$ respectively. If we let $k=\Theta (log
\ n)$, the $n$-node ReCord has $\Theta (\frac{log^2 \ n}{log log \
n })$ node degree and $\Theta (\frac{log \ n}{log log \ n })$
query latency. Fig.~\ref{fig:degree_latency} shows the trade-off
between the node degree and query latency, given the total number
of active nodes is $n=2^{15}=32768$.
Fig.~\ref{fig:degree_latency}, shows that the node degree
increases almost linearly as $k$ increases, but the query latency
drops quickly within a small range of $k$.
%

\section{Implementation of ReCord}
\subsection{Estimation of network size}
Although the analytical results in the previous section can be
directly applied to a static network, the derived bounds for
degree and latency are not as tight as the case using active nodes
in the real P2P network. Now we suppose a dynamic environment,
where the nodes join and leave dynamically. The main difficulty of
this extension is that for each node, it requires a rather
accurate information of the global network size for the
construction of links. When we divide network in each level, we
need to know the value of $n$, the total number of active nodes.

Currently, most estimation processes uses the density information
around the vicinity of the estimating
node~\cite{MNR02,MBR03,HM03}. Let $L_f$ denote the fraction length
of an interval including $f$ distinct nodes. The network size can
be estimated by $\frac{f-1}{L_f}$. In~\cite{MNR02,HM03}, the
length between estimating node and its successor is used to
estimate the size of the overall network. Symphony~\cite{MBR03}
applies the length between estimating node's predecessor and
successor in estimation procedure, and its experiment shows that
the impact of increasing $f$ on average latency is not
significant.

Other methods can be also applied to estimate network size, such
as through a central server, or piggybacking along lookup
messages~\cite{MBR03}, or randomly choosing several pairs of
continuous nodes, and using their average length for estimation.

Knowing the network size is an important step for dynamic network
construction. In our experiments, as with Symphony, we use $f=3$
to estimate the global network size in ReCord.

\subsection{Join and Leave maintenance}
\subsubsection{Join Protocol}

Suppose that a new node $s$ joins the network through an existing
node. Firstly, node $s$ chooses its identifier from $[0,1)$
uniformly at random. Secondly, node $s$ is inserted between
$PRED(s)$ and $SUCC(s)$, and runs the estimation protocol, and
update the estimated network size $\tilde{n_s}$ for all 3 nodes.
Next, it divides the whole ring $[0,1)$  recursively into
intervals $[s+\frac{(j-1)k^i}{ \tilde{n_s}},s+\frac{jk^i}{
\tilde{n_s}})$ starting from $s$ as described in
Section~\ref{sec:construction}. Then it sets up one link to a node
randomly chosen from each interval. The implementation detail for
link construction is that it first generates a random real number
$x$ in the interval $[s+\frac{(j-1)k^i}{
\tilde{n_s}},s+\frac{jk^i}{ \tilde{n_s}})$, then looks up
$SUCC(x)$. If $SUCC(x)$ is in the range $[s+\frac{(j-1)k^i}{
\tilde{n_s}},s+\frac{jk^i}{ \tilde{n_s}})$, the connection is
built successfully, otherwise, it has to re-establish the link for
the interval. In order to avoid flooding traffic made by link
reestablishment, we limit the times of reestablishment. If node
$s$ still can't find a node in an interval after $q$ times tries,
we let it give up the link construction for this interval. The
value of $q$ should be related to the interval length. More
details will be shown in the experiment part.

Similar to Symphony, ReCord also bounds the number of incoming
links per node, which is good for load balancing of the whole
network. Once the number of incoming links of a node has reached
$2log_k \ n $, any new request to establish a link with it will be
rejected. The requesting node has to make another attempt.

Since node $s$ needs a lookup operation that requires $O( \log_k
n)$ messages for each link establishment, the whole cost of $O(k
\log_k  n)$ link constructions is $O(k \log^2_k n)$ messages.

\subsubsection{Leave Protocol}
Once node $s$ leaves the system, all its outgoing links will be
snapped. Its predecessor and successor nodes need to reinstate
their links, and corresponding neighbor nodes need to update their
estimation of network size. At the same time, all the incoming
links of node $s$ are broken, and corresponding connection nodes
need to re-select another node randomly in the same interval as
node $s$ is located in. This operation can be triggered by the
periodic detections by  nodes connected to node $s$.

If node $s$ leaves voluntarily, it will gracefully inform related
nodes to update their connection information, otherwise, the
connection information has to be updated when the other nodes have
periodically detected the failure of node $s$. More details of
this protocol are similar to that in Chord.

\subsection{Re-linking operation}

The total number of active nodes in the P2P network always changes
as the network expands or shrinks. When node $s$ finds that its
current estimated network size $\tilde{n_s}$ is not equal to its
stored stale estimation $\tilde{n'_s}$, it needs to re-build its
links to nodes in the new intervals. One conservative solution is
to re-link every construction on every update of $\tilde{n_s}$. In
this way, the traffic resulted by the re-linking operation would
be excessive. According to analyzed bounds in
Section~\ref{sec:bound_analysis}, the messages required for one
node to re-link all its connections are $O(klog^2_k \ n)$.

In order to avoid the excessive traffic resulted from re-linking
operation, and guarantee the stability of the whole P2P network,
we apply the same re-linking criterion as in Symphony: re-linking
operation occurs only when $\frac{\tilde{n_s}}{\tilde{n'_s}}\geq
2$ or $\frac{\tilde{n_s}}{\tilde{n'_s}}\leq \frac{1}{2}$, where
$\tilde{n_s}$ is node's updated estimated network size, and
$\tilde{n'_s}$ is node's stored stale network size.
%

\subsection{Fault tolerance}

In Chord or Koorde with constant-degree, each node keeps a list of
successors to increase the system's robustness: each node
maintains $r$ connections of its immediate succeeding nodes rather
than only one immediate successor. Certainly, it will keep the
whole P2P network more robust, but it also requires some extra
operations and corresponding additional cost to maintain the
successor list. Using a similar scheme, Symphony makes $r$ copies
of a node's content at each of $r$ succeeding nodes. Other DHTs,
such as CAN, Pastry, and Tapestry keep several backup links for
each node.

Compared with the above DHTs, we found that ReCord has a natural
structure for fault tolerance: at the last dividing level, each
node is already connected to its $k$ following succeeding nodes,
which is equal to a successor list in Chord. ReCord need not keep
any backup link or redundant links to increase the robustness of
the whole system. Therefore, it entails no extra overhead to offer
fault tolerance.

As stated in~\cite{KK03}, in order to keep live nodes connected in
cases of nodes failures, some nodes need to have node degree of at
least $\Omega (log \ n)$. Moreover, the experiments
in~\cite{MBR03} shows that the local links are crucial for
maintaining connectivity of P2P topology. By construction, ReCord
has rich ``local" connections and relatively sparse ``long"
connections. Our experimental results, to be presented shortly,
confirms that this is indeed a good strategy for forming the
connectivity of the P2P network.

\section{Experiments}
Based on the implementation described in the preceding section, we
run the simulation of ReCord with nodes ranging from $2^4$ to
$2^{15}$. The impacts of different parameters shown in the
simulation results are analyzed.

We focus on four types of network. The first one is a
\emph{static} network, in which the global network size is known
for every node. The second one is called an \emph{expanding}
network, where the rate of node joining is higher than that of
node departure in a given unit time interval. The third one,
called a \emph{shrinking} network, is opposite of an expanding
network. The last one is called a \emph{stable} network, in which
the rate of node joining  is equal to that of node departure nodes
in a unit time interval.

\subsection{Estimation of network size}
\begin{figure*}
\centerline{\subfigure[Estimation for small scale
network.]{\includegraphics[width=3.5in]{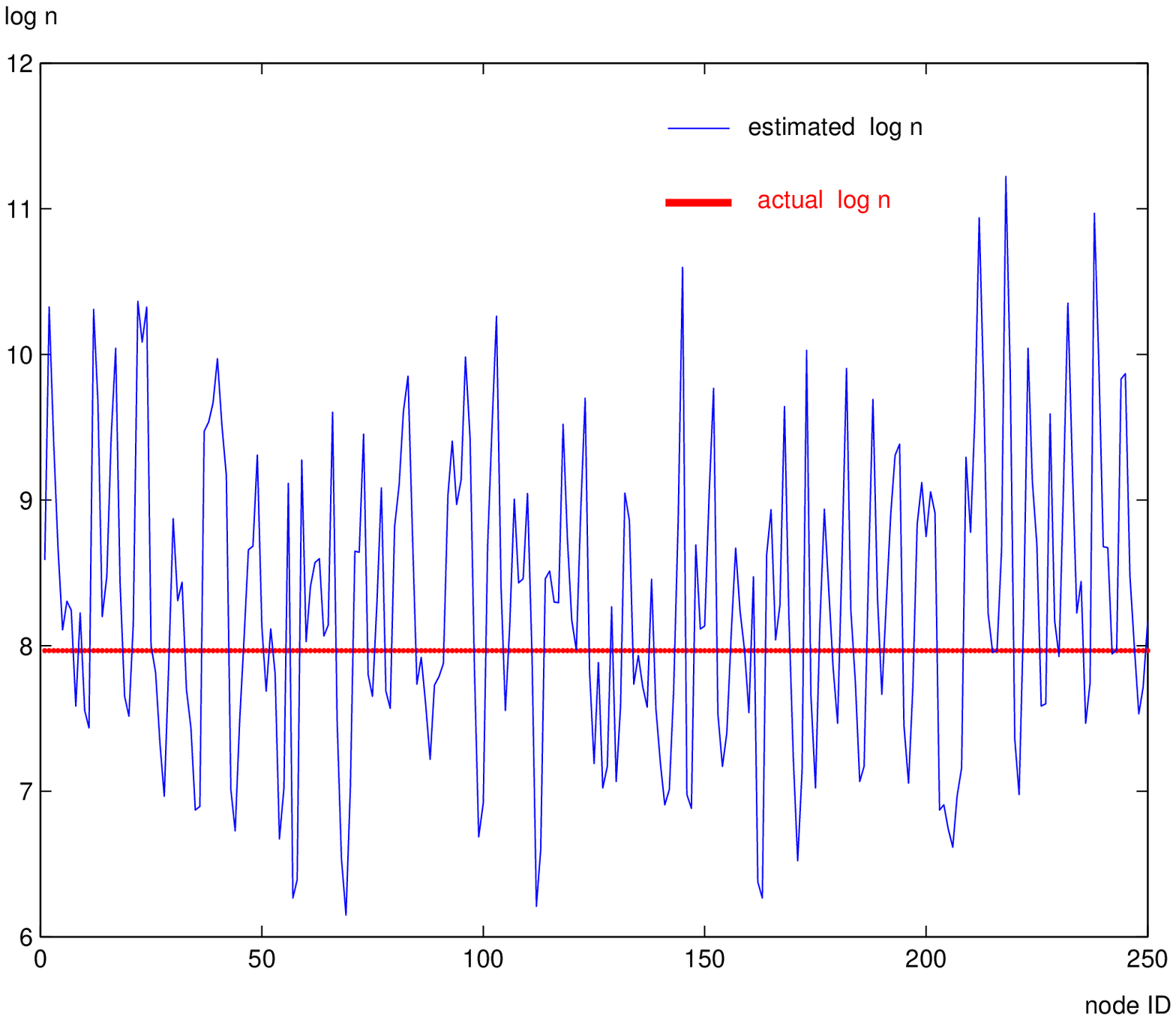}
\label{fig_estimation_smalln_over_nID}} \hfil
\subfigure[Estimation for large scale
network.]{\includegraphics[width=3.5in]{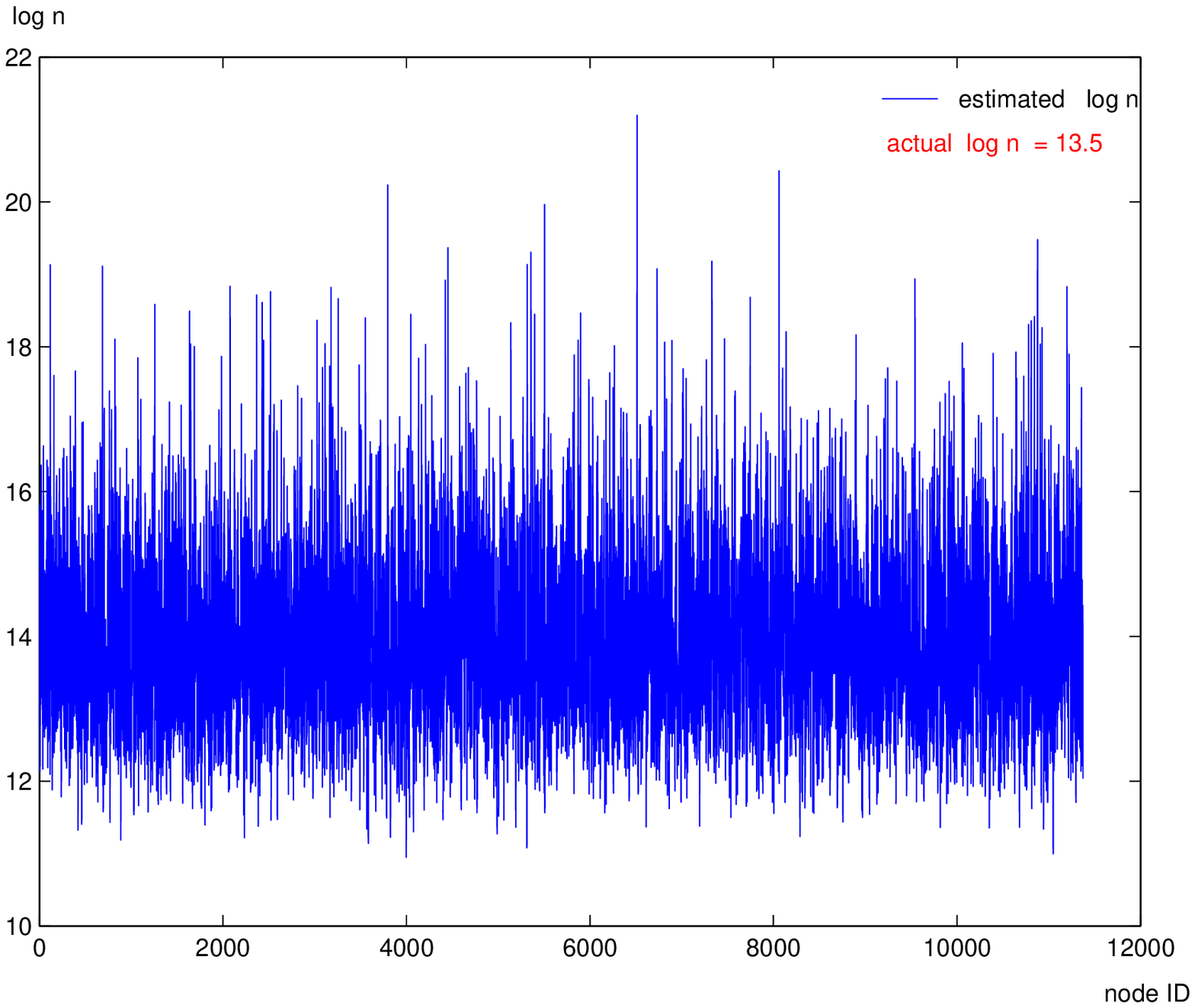}
\label{fig_estimation_largen_over_nID}} }
 \caption{Estimation of network size for stable P2P systems.}
 \label{fig:estimat_n_over_n_ID}
\end{figure*}

For each node, we let it use the density between its predecessor
and successor to estimate the global network size.
Fig.~\ref{fig:estimat_n_over_n_ID} shows the comparisons between
the estimated and the actual network size for both small and large
P2P system which is in the steady state. For the small scale
network with $n=250$ active nodes, the difference between
estimated $log \ \tilde{n}$ and actual $log \ n$ is no more than
4. For a larger scale network with $n=11,374$ active nodes, the
difference between estimated $log \ \tilde{n}$ and actual $log \
n$ is no more than 8. In either network, the difference between
the estimated and actual $log n$ for most nodes is smaller than 4.
This shows that the approximation is accurate enough for the
construction of ReCord.

\begin{figure*}
\centerline{\subfigure[Estimation of network size in expanding
network.]{\includegraphics[width=2.2in]{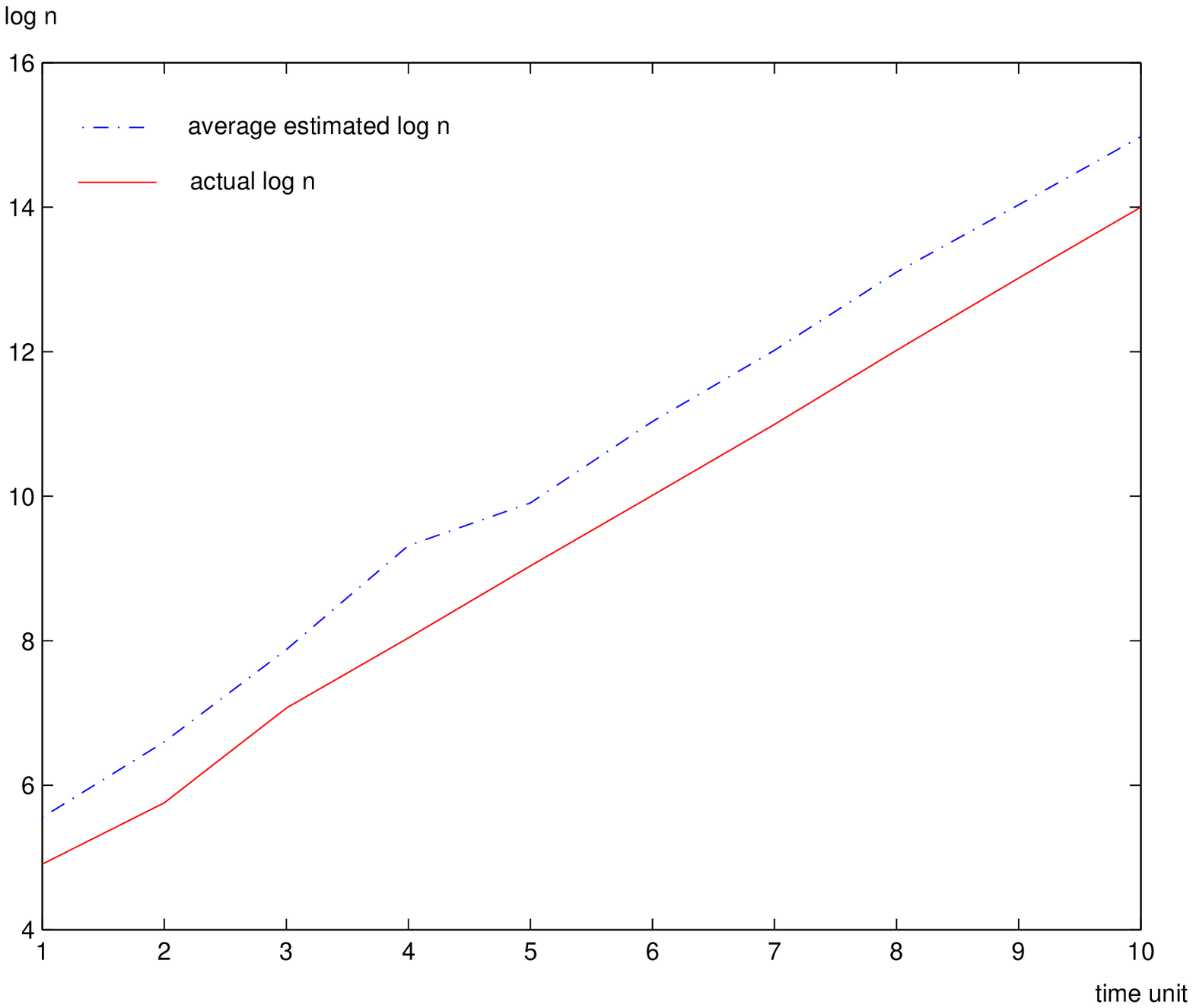}
\label{fig_expanding_estimation_over_time}}
 \hfil \subfigure[Estimation of network size in shrinking
network.]{\includegraphics[width=2.2in]{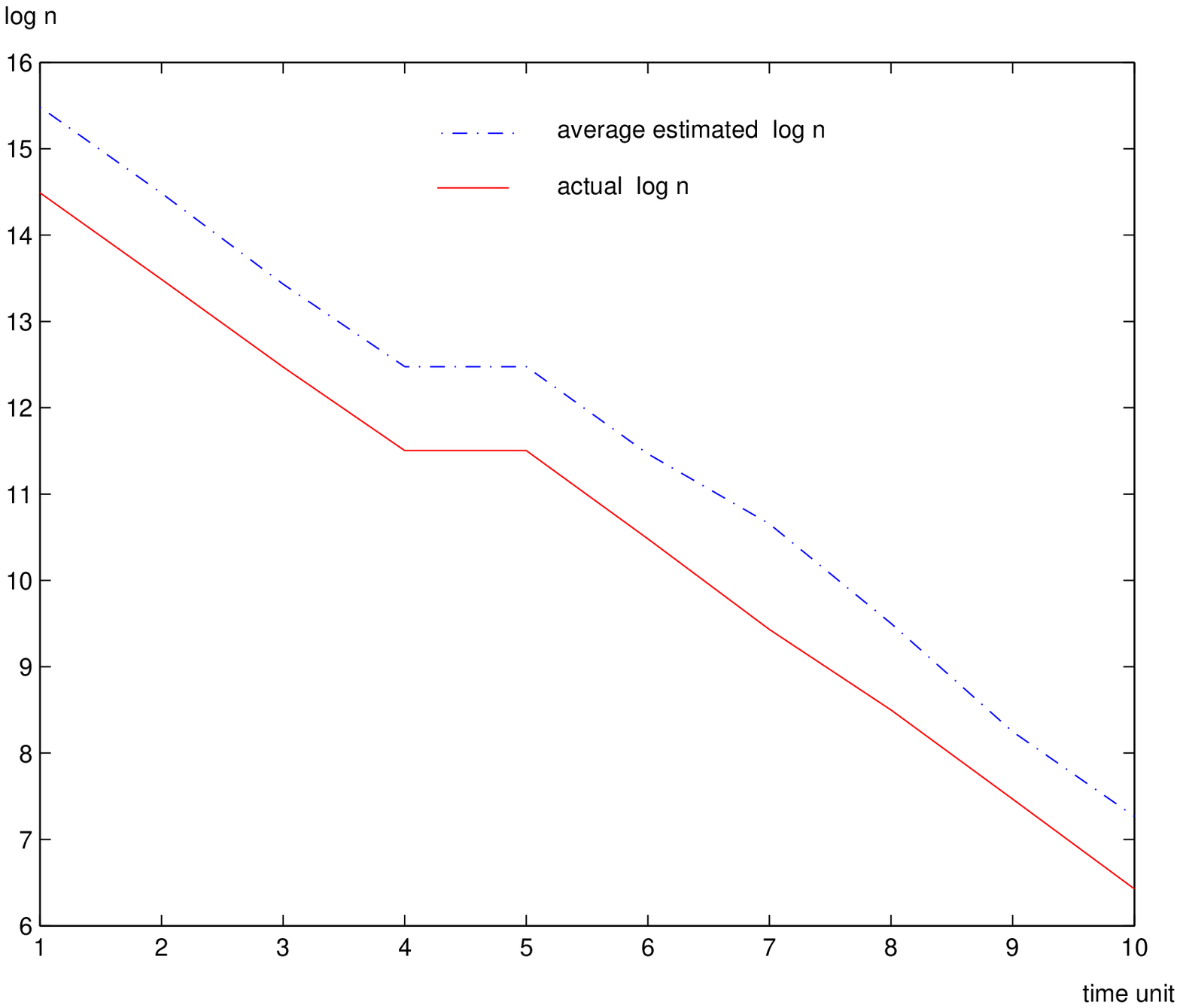}
\label{fig_shrinking_estimation_over_time}} \hfil
\subfigure[Estimation of network size in stable
network.]{\includegraphics[width=2.2in]{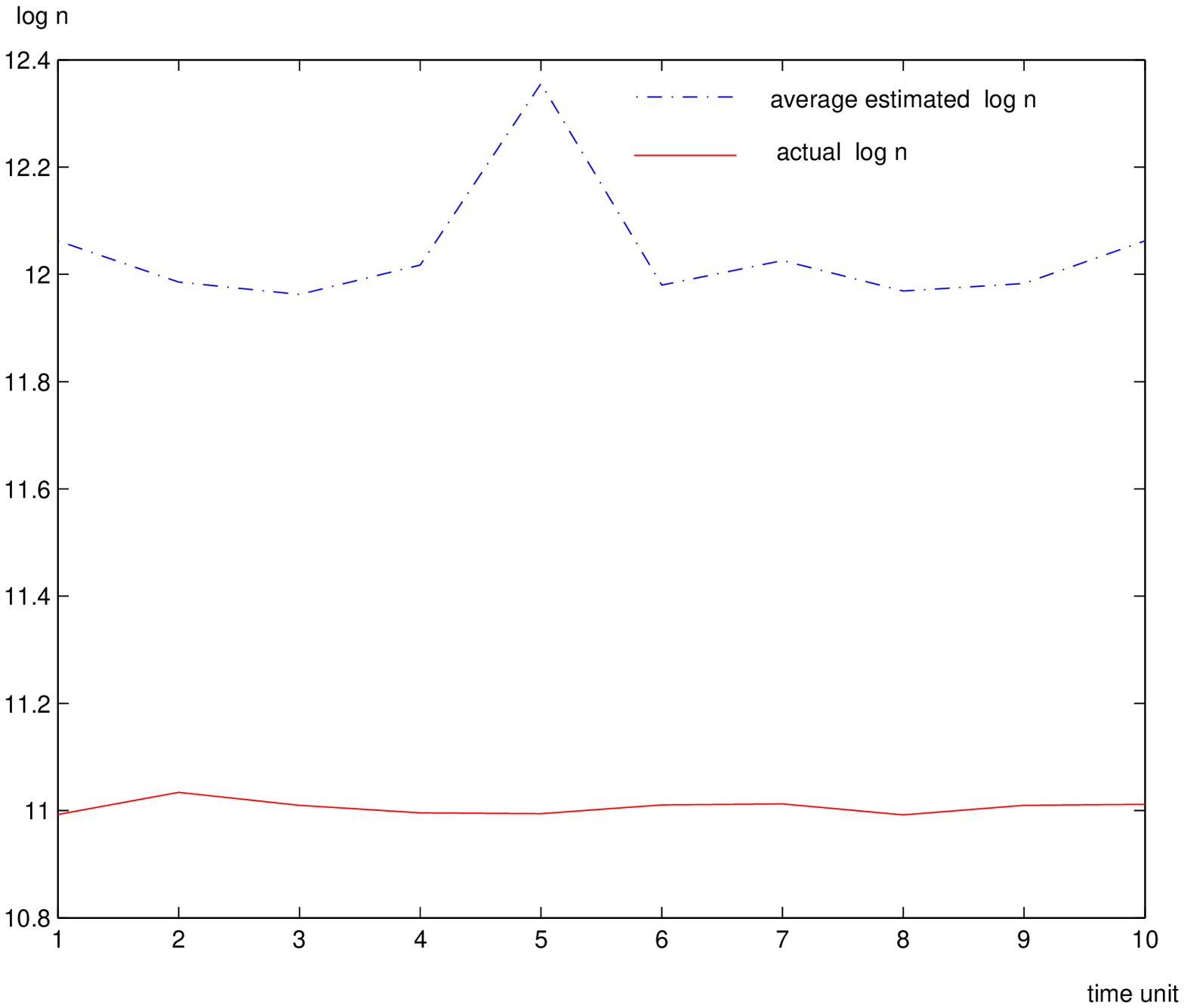}
\label{fig_stable_estimation_over_time}}}
 \caption{Estimation of network size for expanding, shrinking, and stable network respectively.}
 \label{fig:estimation_over_time}
\end{figure*}

 Fig.~\ref{fig:estimation_over_time} shows the average estimation of network size over all
nodes of the expanding, shrinking, and stable networks
respectively over time. The comparisons of average $log \
\tilde{n}$ and actual $log \ n$.

\subsection{Degree and Latency}
\begin{figure*}
\centerline{\subfigure[Average degree for P2P network with $n$
ranging from $2^3$ to
$2^{13}$.]{\includegraphics[width=3.5in]{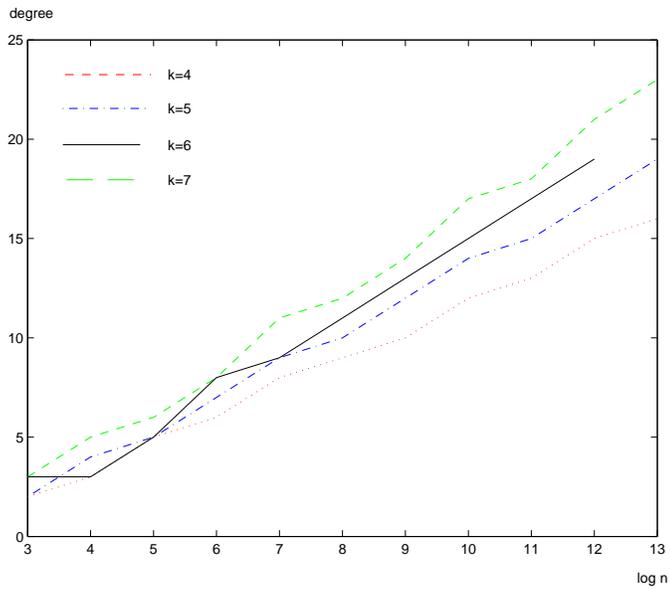}
\label{fig_degree_over_n}} \hfil \subfigure[Average latency for
P2P network with $n$ ranging from $2^3$ to
$2^{13}$.]{\includegraphics[width=3.5in]{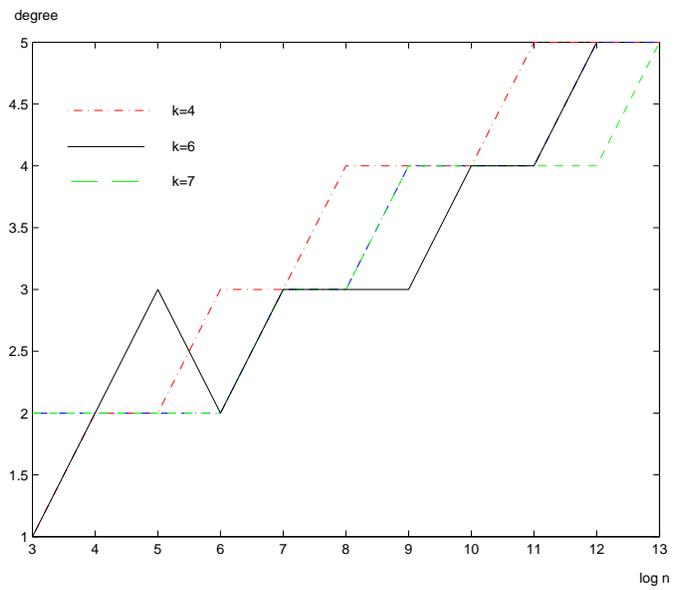}
\label{fig_latency_over_n}} }
 \caption{Average degree and latency on expanding network.}
 \label{fig:degree_lantency_over_n}
\end{figure*}

Fig.~\ref{fig:degree_lantency_over_n} shows the average node
degree and latency over an expanding network with different $k$
values.

\begin{figure}
\centering
\includegraphics[width=3in]{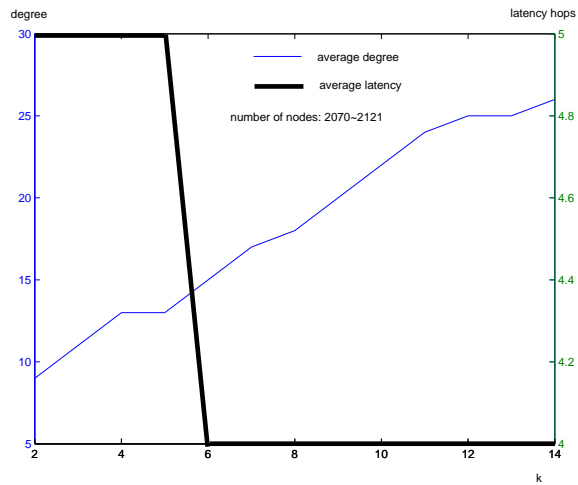}
\caption{Tradeoff between degree and latency over different
choices of $k$ in a stable network (the number of active nodes is
between [2070,2121]).}
\label{fig:degree_latency_given_range_n_over_k}
\end{figure}

Fig.~\ref{fig:degree_latency_given_range_n_over_k} shows the
tradeoff between degree and latency over different $k$ values in a
stable P2P network, given that the number of active nodes is in
the range of $[2070, 2121]$. For
Fig.~\ref{fig:degree_latency_given_range_n_over_k},  $k=5$ is
obviously the best choice for the P2P network of those sizes. We
can also see that it fits the analysis
 ( cf. Fig.~\ref{fig:degree_latency}) quite well.

\subsection{Fault tolerance}

\begin{figure}
\centering
\includegraphics[width=5in]{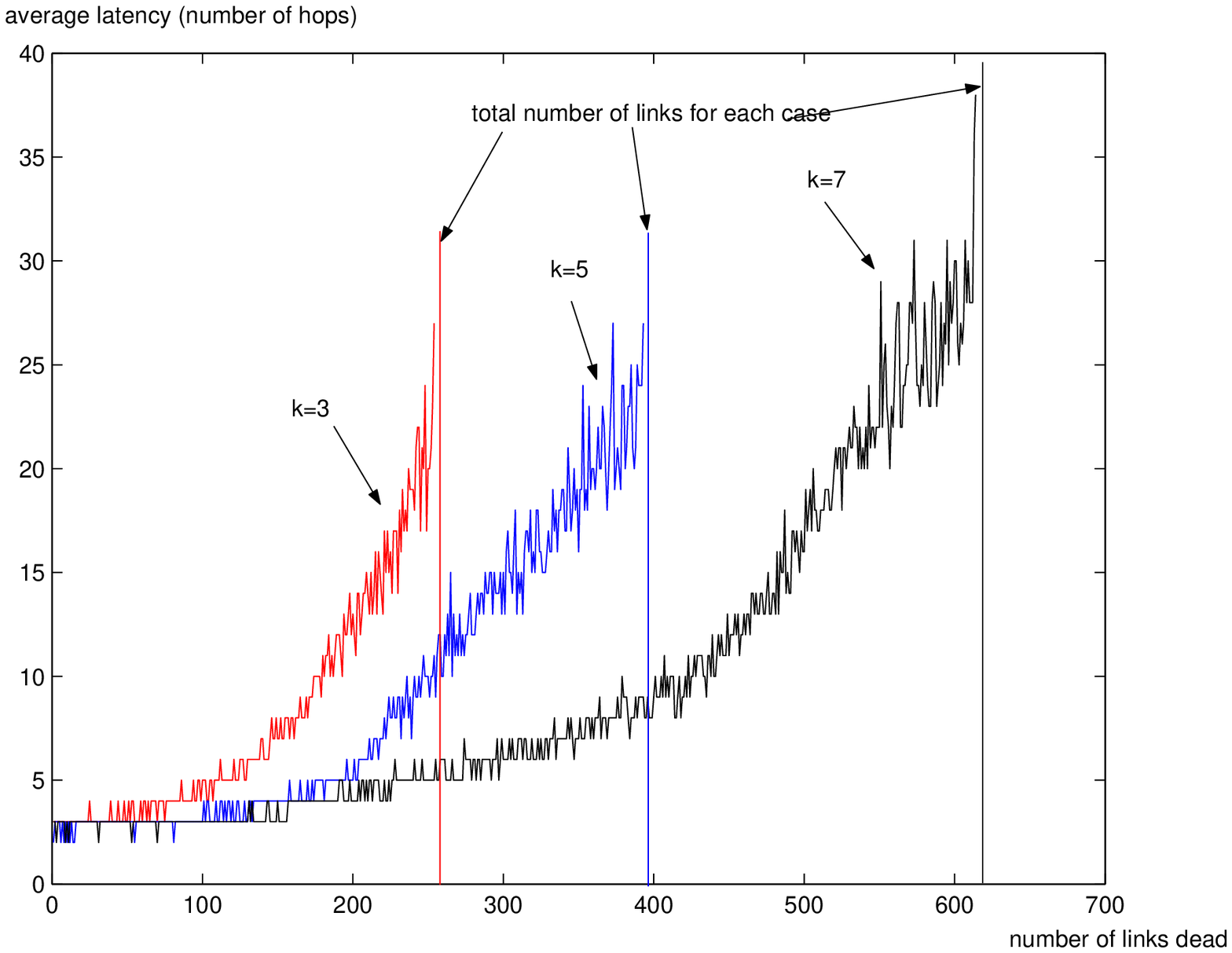}
\caption{Impact of link failures  on the average latency.}
\label{fig:fault_tolerance_log_7_k_357}
\end{figure}

Fig.~\ref{fig:fault_tolerance_log_7_k_357} shows how the fraction
of failure links will influence the query latency. Three cases:
$k=3$, $k=5$, and $k=7$  were run under the stable environment.
According to Fig.~\ref{fig:degree_latency}, the node degree
increases as  $k$ increases for a fixed $n$. However, from
Fig.~\ref{fig:fault_tolerance_log_7_k_357}, we can see that
independent of $k$, only when more than half of the links fail,
the query latency is adversely affected.

\section{Conclusions}

We have proposed a simple DHT topology, called ReCord, which is a
generalization of Randomized-Chord, in the sense that
Randomized-Chord is a special case of ReCord when $k=2$. ReCord
offers a good tradeoff between the node degree and query latency:
an $n$-node ReCord with $O(\log  n)$ node degree has an expected
latency of $\Theta(\log n)$ hops. Alternatively, it can also offer
$\Theta(\frac{\log n}{\log \log n})$ hops latency at a higher cost
of $\Theta(\frac{\log^2  n}{\log \log n})$ node degree. Some
implementation techniques of ReCord are presented, including the
estimation of network size, join and departure maintenance,
re-linking operations, etc. Meanwhile, simulations of the dynamic
behaviors of ReCord are studied.

In actual P2P systems, the different bounds of degree and latency
of the constructed networks usually offer different insights of
DHTs. Lower degree decreases the number of open connections and
the protocol traffic made by the pinging operation; the number of
neighbors whose states need to be changed when a node joins or
leaves is also smaller. However, the lower connectivity of the low
node degree also means that the network is easy to split up, and
hence it has weak fault tolerance. On the other hand, higher node
degree leads to better connectivity and reduces the network
diameter, the longest path length for the query operation. The
lower query latency also leads to lower joining and departure
costs. As we will discuss later, the join and leave operation will
make use of the query operation, so the small path latency will
also decrease the cost of join and leave. We can adjust the $k$
value to suit the P2P networks required for different
environments. A more proactive and perhaps more useful method is
to dynamically monitor the P2P system, e.g. sensing the frequency
of the nodes joining and leaving, and adjusting the $k$ value
dynamically in real time. We will extend our work in this respect
in our future research.

\bibliography{ReCord}

\begin{thebibliography}{10}

\bibitem{BFK01}
L.~Barri\'{e}re, P.~Fraigniaud, E.~Kranakis, and D.~Krizanc.
\newblock Efficient routing in networks with long range contacts.
\newblock In {\em Proceedings of the 15th International Symposium on
  Distributed Computing (DISC'01)}, 2001.

\bibitem{EAB02}
S.~El-Ansary, L.O. Alima, P.~Brand, and S.~Haridi.
\newblock A framework for peer-to-peer lookup services based on $k$-ary search.
\newblock Research Report T2002-06, Department of Microelectronics and
  Information Technology, Royal Institute of Technology, Stockholm, Sweden,
  2002.

\bibitem{GM04optimal}
P.~Ganesan and G.S. Manku.
\newblock Optimal routing in chord.
\newblock In {\em Proceedings of the 15th Annual ACM-SIAM Symposium on Discrete
  Algorithms}, pages 169--178, Jan 2004.

\bibitem{Gnu}
Gnutella.
\newblock http://gnutella.wego.com.

\bibitem{GGG03impact}
K.~Gummadi, R.~Gummadi, S.~Gribble, S.~Ratnasamy, S.~Shenker, and I.~Stoica.
\newblock The impact of dht routing geometry on resilience and proximity.
\newblock In {\em Proceedings of the 2003 conference on Applications,
  technologies, architectures, and protocols for computer communications},
  pages 381--394. ACM Press, 2003.

\bibitem{HM03}
K.~Horowitz and D.~Malkhi.
\newblock Estimating network size from local information.
\newblock {\em Information Processing Letters}, 88(5):237--243, 2003.

\bibitem{KK03}
M.~F. Kaashoek and David~R. Karger.
\newblock Koorde: A simple degree-optimal distributed hash table.
\newblock In {\em Proceedings of the second International Workshop on
  Peer-to-Peer Systems (IPTPS)}, February, 2003.

\bibitem{Kazaa}
Kazaa.
\newblock http://www.kazaa.com.

\bibitem{Kle00}
J.~Kleinberg.
\newblock The {S}mall-{W}orld {P}henomenon: {A}n {A}lgorithmic {P}erspective.
\newblock In {\em Proceedings of the 32nd ACM Symposium on Theory of
  Computing}, 2000.

\bibitem{MNR02}
D.~Malkhi, M.~Naor, and D.~Ratajczak.
\newblock Viceroy: A scalable and dynamic emulation of the butterfly.
\newblock In {\em Proceedings of the 21st {ACM} Symposium on Principles of
  Distributed Computing (PODC 2002)}, 2002.

\bibitem{MBR03}
G.~S. Manku, M.~Bawa, and P.~Raghavan.
\newblock Symphony: Distributed hashing in a small world.
\newblock In {\em Proceedings of the 4th {USENIX} Symposium on Internet
  Technologies and Systems}, 2003.

\bibitem{MNW04}
G.S. Manku, M.~Naor, and U.~Wieder.
\newblock Know thy neighbor's neighbor: The power of lookahead in randomized
  p2p networks.
\newblock In {\em Proceedings of STOC 2004}, 2004.

\bibitem{Napster}
Napster.
\newblock http://www.napster.com.

\bibitem{NZ01towards}
T.~Ng and H.~Zhang.
\newblock Towards global network positioning.

\bibitem{PRR97}
C.~G. Plaxton, R.~Rajaraman, and A.~W. Richa.
\newblock Accessing nearby copies of replicated objects in a distributed
  environment.
\newblock In {\em {ACM} Symposium on Parallel Algorithms and Architectures},
  pages 311--320, 1997.

\bibitem{RFHK01}
S.~Ratnasamy, P.~Francis, M.~Handley, and R.~M. Karp.
\newblock A scalable content-addressable network.
\newblock In {\em Proceedings of ACM SIGCOMM 2001}, pages 161--172, 2001.

\bibitem{RHS02topologically}
S.~Ratnasamy, M.~Handley, R.~Karp, and S.~Shenker.
\newblock Topologically-aware overlay construction and server selection.
\newblock In {\em Proceedings of IEEE INFOCOM'02}, 6 2002.

\bibitem{RST02routing}
Sylvia Ratnasamy, Scott Shenker, and Ion Stoica.
\newblock Routing algorithms for dhts: Some open questions.
\newblock 2002.

\bibitem{RD01}
A.~Rowstron and P.~Druschel.
\newblock Pastry: Scalable, decentralized object location, and routing for
  large-scale peer-to-peer systems.
\newblock {\em Lecture Notes in Computer Science}, 2218:329--350, 2001.

\bibitem{SMKKB01}
I.~Stoica, R.~Morris, D.~Karger, M.~F. Kaashoek, and H.~Balakrishnan.
\newblock Chord: A scalable peer-to-peer lookup service for internet
  applications.
\newblock In {\em Proceedings of the 2001 conference on applications,
  technologies, architectures, and protocols for computer communications},
  pages 149--160. ACM Press, 2001.

\bibitem{ZKJ01}
B.~Y. Zhao, J.~D. Kubiatowicz, and A.~D. Joseph.
\newblock Tapestry: An infrastructure for fault-tolerant wide-area location and
  routing.
\newblock Technical Report UCB/CSD-01-1141, UC Berkeley, April 2001.

\end{thebibliography}

\appendix [Pseudocode for ReCord implemtation] \label{sec:appendix}

\begin{algorithm}
\begin{scriptsize}
\caption{Estimation of network size}
\label{alg:estimation_network_size}
\begin{algorithmic}
\STATE EstimationNetworkSize(s) \ \ // estimate the global network
size for node $s$
\end{algorithmic}
\begin{algorithmic}[1]
 \STATE $length \leftarrow $ the clockwise length between $SUCC(s)$
 and $PRED(s)$; {\ \ //  length of the whole ring is 1}
\STATE $estmated\_number \leftarrow \frac{2}{length}$; \STATE
\textbf{return;}  $estmated\_number$
\end{algorithmic}
\end{scriptsize}\vspace{-3pt}
\end{algorithm}

\begin{algorithm}
\begin{scriptsize}
\caption{Establishment of links for each node}
\label{alg:build_links}
\begin{algorithmic}
\STATE BuildLinks($s,k,length$)\ \ /*build all links for node $s$.
$k$ is the number phases per level. $length$ is the length of
division level. Initially call BuildLinks($s$,$k$,1)*/
\end{algorithmic}
\begin{algorithmic}[1]
\IF {($length/k)<(1/EstimationNetworkSize(s))$} \STATE
\textbf{return}; \ENDIF \FOR{($j=2; j \leq k-1; j++$)} \FOR[
$\sqrt{EstimationNetworkSize(s)*length/k}$ is the bound of retry
times ] {($retry\_time=1; retry\_time \leq
\sqrt{EstimationNetworkSize(s)*length/k}; retry\_time++$)} \STATE
$rand \leftarrow $ a random number between [0,1); \STATE $temp\_n
\leftarrow$ $FindSuccessor(temp\_n,s)$; \IF
[$2*k*log(EstimationNetworkSize(s))$ is the limitation of incoming
degree per node]{incoming degree of node $temp_n \leq
2*k*log(EstimationNetworkSize(s))$ } \IF [$ID(s)$ the identifier
of node after mapped into $[0,1)$ ]{$rand \in [ID(s)+
\frac{j*(length/k)}{EstimationNetworkSize(s)},ID(s)+
\frac{(j+1)*(length/k)}{EstimationNetworkSize(s)})$}
 \STATE build a connection between node $s$ and node $temp\_n$; \ENDIF \ENDIF
  \ENDFOR \ENDFOR
\STATE BuildLinks(s,k,length/k)
\end{algorithmic}
\end{scriptsize}\vspace{-3pt}
\end{algorithm}

\begin{algorithm}
\begin{scriptsize}
\caption{Find the successor for a given identifier}
\label{alg:find_successor}
\begin{algorithmic}
\STATE FindSuccessor($s,x$)\ \ /*find the successor of identifier
$x$ through node $s$*/
\end{algorithmic}
\begin{algorithmic}[1]
\IF [$ID(s)$ is the identifier of node $s$ mapped into $[0,1)$]{$x
\in [ID(s),ID(s+1))$} \STATE \textbf{return} $s+1$; \ELSE \STATE
{\textbf{return} $FindSuccessor(s+1,x)$} \ENDIF
\end{algorithmic}
\end{scriptsize}\vspace{-3pt}
\end{algorithm}

\begin{algorithm}
\begin{scriptsize}
\caption{Find the predecessor for a given identifier}
\label{alg:find_predecessor}
\begin{algorithmic}
\STATE FindPredecessor(s,x)\ \ /*find the predecessor of
identifier $x$ through node $s$*/
\end{algorithmic}
\begin{algorithmic}[1]
\IF [$ID(s)$ is the identifier of node $s$ mapped into $[0,1)$]{$x
\in [ID(s),ID(s-1))$} \STATE \textbf{return} $s-1$; \ELSE \STATE
{\textbf{return} $FindSuccessor(s-1,x)$} \ENDIF
\end{algorithmic}
\end{scriptsize}\vspace{-3pt}
\end{algorithm}

\begin{algorithm}
\begin{scriptsize}
\caption{Join operation} \label{alg:join}
\begin{algorithmic}
\STATE Join($n_0$,$x$)\ \ /*node $s$ joins the system through node
$n_0$*/
\end{algorithmic}
\begin{algorithmic}[1]
\STATE $n_1 \leftarrow FindPredecessor(n_0,s)$; \STATE $SUCC(n_1)
\leftarrow s$; \STATE $PRED(s) \leftarrow n_1$;

\STATE $n_2 \leftarrow FindSuccessor(n_0,s)$; \STATE $SUCC(s)
\leftarrow n_2$; \STATE $PRED(n_2) \leftarrow s$;

\STATE EstimationNetworkSize($s$)

\STATE BuildLinks($s$,$k$,1)

\end{algorithmic}
\end{scriptsize}\vspace{-3pt}
\end{algorithm}
\begin{algorithm}
\begin{scriptsize}
\caption{Leave operation} \label{alg:Leave}
\begin{algorithmic}
\STATE Leave($x$)\ \
\end{algorithmic}
\begin{algorithmic}[1]
\STATE $SUCC(PRED(s)) \leftarrow SUCC(s)$; \STATE $PRED(SUCC(s))
\leftarrow PRED(s)$; \STATE delete all incoming and outcoming
links of node $s$ and inform corresponding nodes;
\end{algorithmic}
\end{scriptsize}\vspace{-3pt}
\end{algorithm}
\begin{algorithm}
\begin{scriptsize}
\caption{Lookup operation} \label{alg:lookup}
\begin{algorithmic}
\STATE Lookup($x$)\ \ /* lookup identifier $x$ through node $s$ */
\end{algorithmic}
\begin{algorithmic}[1]
\STATE $min\_length \leftarrow 0$; \STATE $min\_neighbor
\leftarrow SUCC(s)$;

\IF {FindSucessor($s$,$x$)=$s$} \STATE \textbf{return} 0; \ENDIF

\STATE $min\_length \leftarrow$ clockwise distance between
$min\_neighbor$ and $FindSuccessor(x)$;

\FOR {$n_i \in$ neighbors of nodes} \IF { clockwise distance
between $n_i$ and $FindSuccessor(x)$<$min\_length$} \STATE
$min\_length \leftarrow$ clockwise distance between $n_i$ and
$FindSuccessor(x)$; \STATE $min\_neighbor \leftarrow n_i$;
 \ENDIF
\ENDFOR \STATE \textbf{return} Lookup($min\_neighbor$,$x$)+1;
\end{algorithmic}
\end{scriptsize}\vspace{-3pt}
\end{algorithm}

\begin{algorithm}
\begin{scriptsize}
\caption{Relinking operation} \label{alg:relink}
\begin{algorithmic}
\STATE Relink($s$)\ \
\end{algorithmic}
\begin{algorithmic}[1]
\IF[$\tilde{n}$ denotes new estimated $n$; $\tilde{n'}$ denotes
old estimated $n$ ] {$\frac{\tilde{n}}{\tilde{n'}}<0.5$  or
$\frac{\tilde{n}}{\tilde{n'}}>2$}
 \STATE delete all outcoming links of node $s$ and inform corresponding
nodes;

\STATE BuildLinks(s,k,1)
 \ENDIF
\end{algorithmic}
\end{scriptsize}\vspace{-3pt}
\end{algorithm}
\end{document}